\documentclass[showpacs,aps,11pt]{revtex4}

\usepackage{epsfig}
\usepackage{graphicx}
\usepackage{amsmath,amssymb,color}
\usepackage[english]{babel}
\usepackage{natbib}
\usepackage{amsmath}
\usepackage{amsfonts}
\usepackage{amssymb}
\usepackage{epstopdf}
\usepackage{enumerate}

\parskip=\medskipamount

\newcommand\disp{\displaystyle}

\newcommand{\eq}[1]{(\ref{#1})}
\newcommand{\fig}[1]{Fig.\ref{#1}}

\newcommand{\be}{\begin{equation}}
\newcommand{\ee}{\end{equation}}
\newcommand{\bea}{\begin{eqnarray}}
\newcommand{\eea}{\end{eqnarray}}

\newcommand{\la}{\left<}
\newcommand{\ra}{\right>}

\begin{document}

\title{From statistics of regular tree-like graphs to
distribution function and gyration radius of branched polymers}

\author{Alexander Y. Grosberg$^{1,2}$, and Sergei K. Nechaev$^{3,4}$}

\affiliation{$^1$Physico-Chimie Curie UMR 168, Institut Curie, PSL Research University, 26 rue
d'Ulm, 75248 Paris Cedex 05, France \\ $^2$Department of Physics and Center for Soft Matter
Research, New York University, 4 Washington Place, New York, NY 10003, USA \\
$^3$Universit\'e Paris--Sud/CNRS, LPTMS, UMR8626, B\^at. 100, 91405 Orsay, France \\
$^4$P.N. Lebedev Physical Institute of the Russian Academy of Sciences, 119991, Moscow, Russia}

\date{\today}

\begin{abstract}

We consider flexible branched polymer, with quenched branch structure, and show that its
conformational entropy as a function of its gyration radius $R$, at large $R$, obeys, in the
scaling sense, $\Delta S \sim R^2/(a^2L)$, with $a$ bond length (or Kuhn segment) and $L$ defined
as an average spanning distance.  We show that this estimate is valid up to at most the logarithmic
correction for any tree.  We do so by explicitly computing the largest eigenvalues of Kramers
matrices for both regular and ``sparse'' 3-branched trees, uncovering on the way their peculiar
mathematical properties.

\end{abstract}


\maketitle

\section{Introduction}

We are interested in statistical mechanics of long flexible branched objects. The leading example,
but certainly not the only one, is large RNA molecule. By means of covalent bonds acting between
nucleotides along a chain and non-covalent saturating bonds between complementary bases, RNA
molecule folds in a peculiar secondary structure which is effectively a branched polymer. There is
an enormous literature on RNA, and on branched polymers in general, however we will be interested
in a specific question which has not been sufficiently considered so far: what is the
conformational entropy of a branched polymer in a tree-dimensional space, and how does it depend on
the overall spatial span (e.g., on the gyration radius), and on the specific arrangement of
branches.  To make it clear: we imagine a branched polymer, such as a secondary structured RNA, to
flex its bonds and junctions without re-arranging secondary structure itself, which we will refer
to as \textit{quenched} branched polymer; we will be interested in entropy associated with spatial
conformations of this quenched object.

Conformational entropy of this type was first written down by Daoud {\it et al}
\cite{DaoudPincusStockmayerWitten_1983}, as $\Delta F/T \sim R^2/R_0^2 $, where $R$ and $R_0$ have
been denoted as ``actual'' and ``unperturbed'' gyration radii, respectively. In the work
\cite{DaoudPincusStockmayerWitten_1983}, authors examined the influence of excluded volume
interactions on the swelling of branched polymers and on the properties of branched polymers in the
solutions, so for them $R$ was the average size of a self-avoiding polymer, while $R_0$ was
considered as ideal. Actually, $R_0$ was taken as $R_0 \sim a N^{1/4}$, i.e., as a mean size of an
ideal tree \cite{Zimm_Stockmayer}, where averaging was supposed to include both spatial
fluctuations of a given tree, and all rearrangements of the tree branches (i.e., it was considered
an \textit{annealed} tree). It is this double average that leads to the well known Zimm-Stockmayer
law, $R_0 \sim a N^{1/4}$, for ideal branched polymers \cite{Zimm_Stockmayer}.

Later on, a similar looking expression,
\be \Delta F/T \sim R^2/(a^2 L) \ ,
\label{F_in_terms_of_L}
\ee
was employed in a more subtle context, emphasizing the difference between quenched and annealed
branched polymers \cite{Annealed_Branched}. In this case, $a^2 L$ replaces $R_0^2$. Obviously, in
this sense, $L$ can be thought as a length of a Gaussian polymer of a typical mean squared size
$\sim R_0^2$. More meaningfully, $L$ was interpreted as an order parameter, the average spanning
distance of a tree. This theory (applied in a variety of contexts \cite{Rosa_Everaers_PRL_2014,
Grosberg_ring_melt_static, Smrek_Grosberg_annealed_trees_dynamics_2014, Bruinsma_virus}) implies
the necessity to consider conformational entropy of a branched polymer with fixed (quenched)
branches, characterized by $L$. From this point of view, the free energy estimate $\Delta F/T \sim
R^2/(a^2L)$ may seem suspicious.  Indeed, if $L$ is the spanning distance of the tree, we can
imagine selecting one particular line of exactly $L$ bonds in the tree (call it a tree trunk), and
then $R^2/(a^2L)$ is (or seems to be?) the free energy of this tree trunk viewed as a linear
polymer.  In fact, to relate this to the free energy of a tree, two aspects have to be taken into
account: first, when the tree is stretched (or swollen), then all its branches are stretched, not
only the trunk; second, if $R$ is the gyration radius of the tree, then gyration radius of the
trunk is somewhat smaller than $R$ (because branches provide a lot of mass close to the center).
This leads us to the question: Is the free energy estimate, $\Delta F/T \sim R^2/(a^2L)$ valid for
every tree, with $L$ defined as an average spanning distance, at least in the scaling sense? In
this paper, we set out to address this question, and our answer is positive: we will show that this
estimate of entropy is valid for any tree, up to at most the logarithmic correction.

\begin{figure}[ht]
\epsfig{file=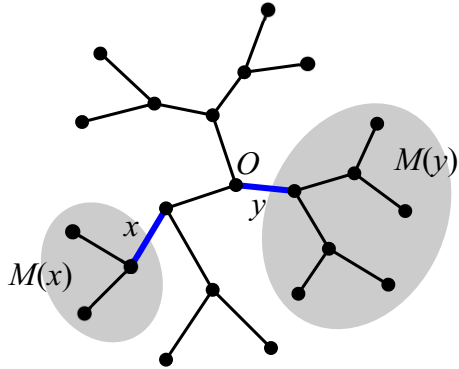,width=6cm}
\caption{Towards the definition of matrix $G_{x,y}$. On the tree, the bonds $x$ and $y$ naturally
divide all nodes of the graph into three categories: on one side of $x$ (shaded), on the other side
of $y$ (also shaded), and in between (not shaded).  We denote the numbers of monomers in the former
two categories as $M(x)$ and $M(y)$. Each matrix element is given by $G_{x,y}=M(x)M(y)/N^{2}$. In
this particular example $G_{x,y}=3\times 7/22^{2}$.}
\label{fig:Gmatrix}
\end{figure}

We build our arguments based on the so-called Kramers theorem \cite{Kramers} (see also \cite{Rubinstein_book}) and its recent
generalization found in \cite{Smrek_Grosberg_annealed_trees_dynamics_2014}. These statements can be
formulated as follows. Let us label all the bonds of the tree (in arbitrary order) with the index
$x$, $1 \leq x \leq N-1$, where $N$ is the number of monomers (tree vertices). For a Gaussian
system (without excluded volume, and when ``bonds'' are long enough compared to the persistence
length), each bond vector, $\mathbf{a}$, is normally distributed, $\sim e^{-3 \mathbf{a}^2 /2
a^2}$, with the mean squared bond length $a^2 = \la \mathbf{a}^2 \ra$, where the vectors
$\mathbf{a}_x$ and $\mathbf{a}_y$ are independent for any $x \neq y$. Let us further define the
$(N-1) \times (N-1)$ matrix $G$ with the entries $G_{x,y}=M(x)M(y)/N^2$, where $M(x)$ and $M(y)$
are the numbers of tree vertices on the one side of bond $x$, while $M(y)$ is the number of
vertices on the other side of bond $y$ -- see the illustration in the \fig{fig:Gmatrix} and more
detailed discussion in the Section II. In terms of this matrix the Kramers theorem reads:

\begin{description}
\item[Kramers theorem.] The averaged gyration radius of the tree is given by the trace ${\rm tr}\,
G$:
\be
\la R^{2}/a^2 \ra = \sum_{x} G_{x,x} = \frac{1}{N^2} \sum_{x=1}^{N-1} M(x)(N-M(x)) \ .
\label{Gkk}
\ee
Here and below, $\la \ldots \ra$ means quenched average, i.e., average over all spatial
conformations with fixed tree structure.

\item[Generalization of Kramers theorem.] For $R \gg \la R \ra$, the probability $P(R)$ of a
given branched polymer (with quenched tree structure) to have a gyration radius $R$, up to power
law corrections, goes as
\be
\left. P(R) \right|_{R \to \infty} \sim e^{- 3 R^2/(2 a^2 \lambda_{\max})} \ ,
\label{eq:Kramers_gen}
\ee
where $\lambda_{\max}$ is the largest eigenvalue of the matrix $G$. In other words, the free energy
price for swelling a quenched tree to a large size $R$, up to logarithmic corrections, goes as
\be
\Delta F/T \simeq 3R^2/(2a^2 \lambda_{\max}) \ .
\label{eq:Kramers2}
\ee

\end{description}

The former statement (the Kramers theorem itself) can be also found in the textbook
\cite{Rubinstein_book}, while the latter statement (its generalization) was proven recently in
\cite{Smrek_Grosberg_annealed_trees_dynamics_2014}. To make the present work self-contained, we
reproduce the derivation in the Appendix \ref{sec:Kramers}.

The result \eq{eq:Kramers2} of the generalized Kramers theorem does not explain the relation
between $\lambda_{\mathrm{max}}$ and the internal geometry/topology of the tree itself.  In this
paper we will establish the following:

\begin{itemize}

\item For a regularly $p=3$-branching tree (a ``3-dendrimer'') $\lambda_{\max} =c_{\rm tree}$ with
$c_{\rm tree} \approx 0.957$;

\item For a regular sparse tree, where each bond is a linear polymer of length $s$,
$\lambda_{\max} = s c_{\rm tree}$;

\item For a linear polymer (or palm tree, with trunk, but without branches), $\lambda_{\max}= N/\pi^2$;

\item We provide an interpolation of $\lambda_{\max}$ from ``dense'' ($s=1$) to ``sparse''
$N> s \gg 1$ (almost palm) trees in the form $\lambda_{\max} \simeq  \overline{c} s$, where
$\overline{c}$ is another constant of order of unity.

\item Based on the above examples, we conclude that in general, for any tree,
$\lambda_{\mathrm{max}} = \tilde{c} L$, where $L$ is the average spanning distance of the tree,
and factor $\tilde{c}$ is at most of the order of $\ln N$.

\end{itemize}

The last statement makes the bridge between equations (\ref{F_in_terms_of_L}) and
(\ref{eq:Kramers2}), thus proving the former.

\section{Largest eigenvalue of Kramers matrix for regular trees}

\subsection{Trees with branchings at every node}

\begin{figure}[ht]
\epsfig{file=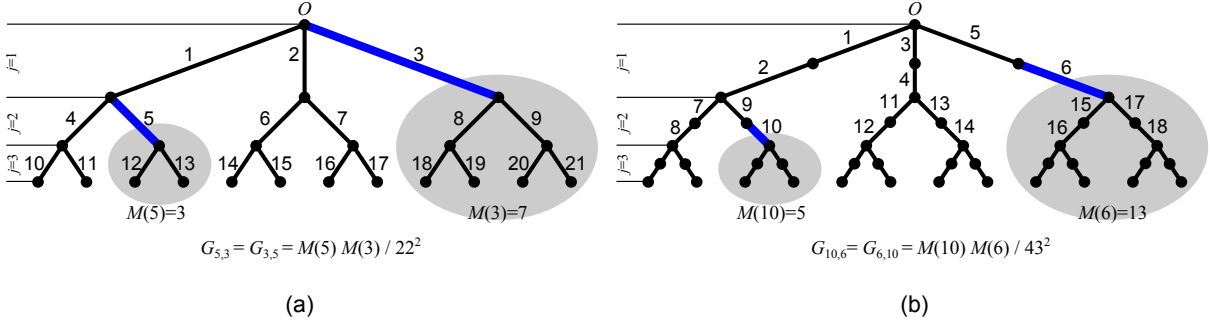,width=16cm}
\caption{(a) The example of the 3-branching finite tree of $k=3$ levels (generations). Vertices
are enumerated by natural numbers sequentially left-to-right in descending order. The shadowed
areas mark the subtrees associated with links $x=5$ and $x=3$. The mass of the subtree, $M(x)$, is
counted in the number of vertices in this subtree. The matrix element, $G_{x,y}$ of the adjacency
matrix $G$ is the product of masses for links $x$ and $y$; (b) An example of ``sparse'' 3-branching
tree of $k=3$ levels. Everything is the same as in (2), however each branching point of the sparse
tree is separated by the subchain of $s$ (in the figure (b) $s=2$) links. The tree links are
enumerated sequentially first in linear subchains and then -- as in (a).}
\label{fig:01}
\end{figure}

To get an explicit form provided by the estimate \eq{eq:Kramers2}, we computed the highest
eigenvalue of adjacency matrix $G$ of a regular symmetric tree. For simplicity we consider
$3$-branching trees only, however results can be easily generalized for trees with any branching
number, $p$. To specify the system, suppose that the tree has $k$ generations, and the total number
of vertices is $N_k$. It is convenient to represent a tree by descending levels (generations) as
shown in the \fig{fig:01}. The distance from the root point, $O$, is counted in the number of
levels. In order to construct the adjacency matrix $G$ corresponding to a tree, we should enumerate
somehow the tree links. Apparently, the most straightforward and naive is the sequential
enumeration of links in each level. Namely, we start with a left-most link in the 1st level and
enumerate all links in this generation left-to-the-right, then proceed in the same way with the
next level, etc. Thus, all links are enumerated sequentially by natural number $x$, where
$x=1,2,..., 3\times 2^{k-1}$.

\begin{figure}[ht]
\epsfig{file=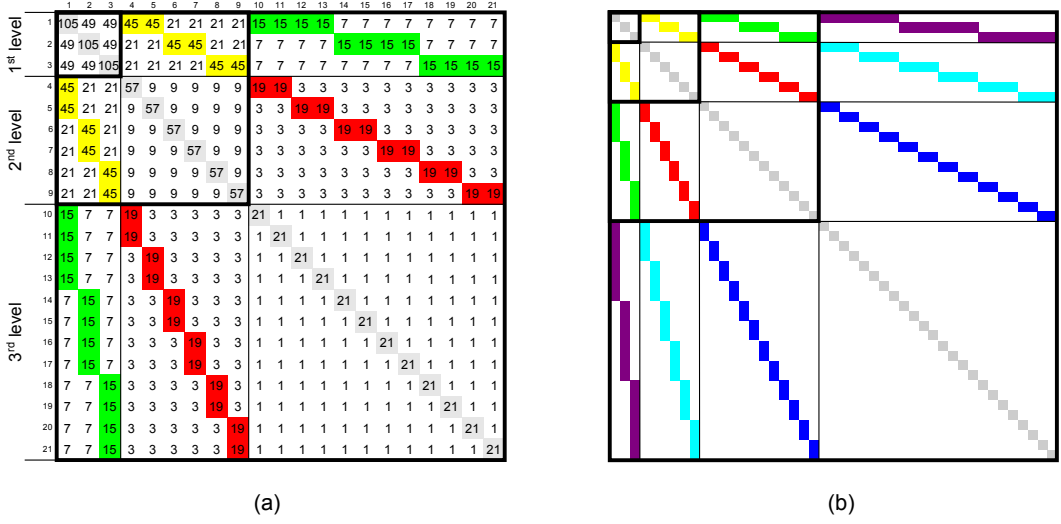,width=14cm}
\caption{(a) The example of the adjacency matrix $\tilde{G}$ corresponding to the tree in
the \fig{fig:01}. The matrix $\tilde{G}$ clearly displays a hierarchical structure. By colors we
highlight the principal repetitive sequences; (b) The generic structure of the adjacency matrix
$\tilde{G}$.}
\label{fig:02}
\end{figure}

Let us repeat the construction of the adjacency matrix $G=\{G_{x,y}\}$ of the regular tree in the
descent diagram representation (\fig{fig:01}). Define the ``mass,'' $M(x)$, associated with the link
$x$ as the sum of all vertices in the subtree which begins with the link $x$. The matrix element
$G_{x,y}=G_{y,x}$ is the non-normalized product of masses $M(x)$ and $M(y)$ for links $x$ and $y$
correspondingly: $G_{x,y}=M(x)M(y)/N_3^2$. In the \fig{fig:01} we have demonstrated the
construction of the element $G_{3,5}=G_{5,3}$ for the adjacency matrix of size $N_3\times
N_3=22\times 22$. The explicit form of the matrix $G$ for the tree shown in the \fig{fig:01}a, is
given in the \fig{fig:02}a. It is instructive to have a dictionary with necessary definitions:

\begin{itemize}

\item The distance from the root point on the tree, counted in the number of levels, is denoted by
$j$, the total number of levels in the tree is $k$;

\item The total number of vertices, $N_k$, of the $k$-level 3-branching tree is $N_k=3\times 2^k-2$;

\item The number of vertices, $N_j$, in the subtree starting from some link in the level $j$, is
$N_{j,k} = 2^{k-j+1}-1$, where $1\ge j\le k$. (In the \fig{fig:01}a the subtrees with
$N_{j=2,k=3}=3$ and $N_{j=1,k=3}=7$ are shown). The ``mass,'' $M$, of the subtree is counted in the
number of vertices of this subtree.

\end{itemize}

The generic structure of the square $N_k\times N_k$ adjacency matrix $\tilde{G}$ with components
$\tilde{G}_{x,y}=N^2 G_{x,y} \equiv M(x)M(y)$, (where $N_k=3\times 2^k-2$), is schematically shown
in the \fig{fig:02}b.

Let us denote by ${\bf U}^{(s)}=\left\{U_1^{(s)}, ..., U_{N_k}^{(s)}\right\}$ the eigenvector $s$
of the matrix $G$ corresponding to the eigenvalue $\lambda_s$ ($1\le s\le N_k$), some of
eigenvalues and eigenvectors may be degenerated. We noticed that due to the symmetry properties of
the adjacency matrix $G$, its leading (maximal) eigenvector $U^{\max}$ has the form
\be
{\bf U}^{\max}=\Big\{\overbrace{U_1^{\max},\;U_1^{\max},\; U_1^{\max}}^{\rm level\; 1\; (3\;
elements)};\; \overbrace{U_2^{\max},\; ...,\; U_2^{\max}}^{\rm level\; 2\; (3\times 2\;
elements)};\; ...;\; \overbrace{U_k^{\max},\; ...,\; U_k^{\max}}^{{\rm level}\; k\; (3\times
2^{k-1}\; {\rm elements})}\Big\}
\label{eq:02}
\ee
Plugging this ansatz into the equation $G U^{\max} = \lambda_{\max} U^{\max}$, we find the leading
eigenvalue $\lambda_{\max}$ to be also the leading eigenvalue of an exponentially smaller matrix
$B$:
\be
B\,{\bf V}^{\max} = \lambda_{\max}{\bf V}^{\max}
\label{eq:03}
\ee
where ${\bf V}^{\max}=\{U_1^{\max},\, U_2^{\max},\,...,\,U_k^{\max}\}$, and the matrix $B$ reads
\be
B=\left(\begin{array}{rrrrrr}
b_{11} & b_{12} & b_{13} & b_{14} &  \ldots & b_{1k} \medskip \\
\disp \frac{b_{12}}{2}  & b_{22} & b_{23} & b_{24} &  \ldots & b_{2k} \medskip \\
\disp \frac{b_{13}}{4} & \disp \frac{b_{23}}{2} & b_{33} & b_{34}  & \ldots & b_{3k} \medskip \\
\disp \frac{b_{14}}{8} & \disp \frac{b_{24}}{4} & \disp \frac{b_{34}}{2} & b_{44} &  &
b_{4k} \medskip \\ \vdots & \vdots & \vdots &  & \ddots & \medskip \\
\disp \frac{b_{1k}}{2^{k-1}} & \disp \frac{b_{2k}}{2^{k-2}} & \disp \frac{b_{3k}}{2^{k-3}} & \disp
\frac{b_{4k}}{2^{k-4}} & & b_{kk}
\end{array}\right); \quad
\begin{array}{ll}
\disp b_{ij} = & 2^{-1 - 2 i}(2^{3 + k} - 3\times 2^{2 + i + k} + 3\times 2^{2i}) \\ & \disp \times
\frac{2^j - 2^{1 + k}}{N_k^2} \qquad (j\ge i)
\end{array}
\label{eq:04}
\ee
(recall that $N_k=3\times 2^k-2$). Unlike $G$, which is $N_k \times N_k$, i.e., exponentially
large, the matrix $B$ is only $k \times k$.

\begin{figure}[ht]
\epsfig{file=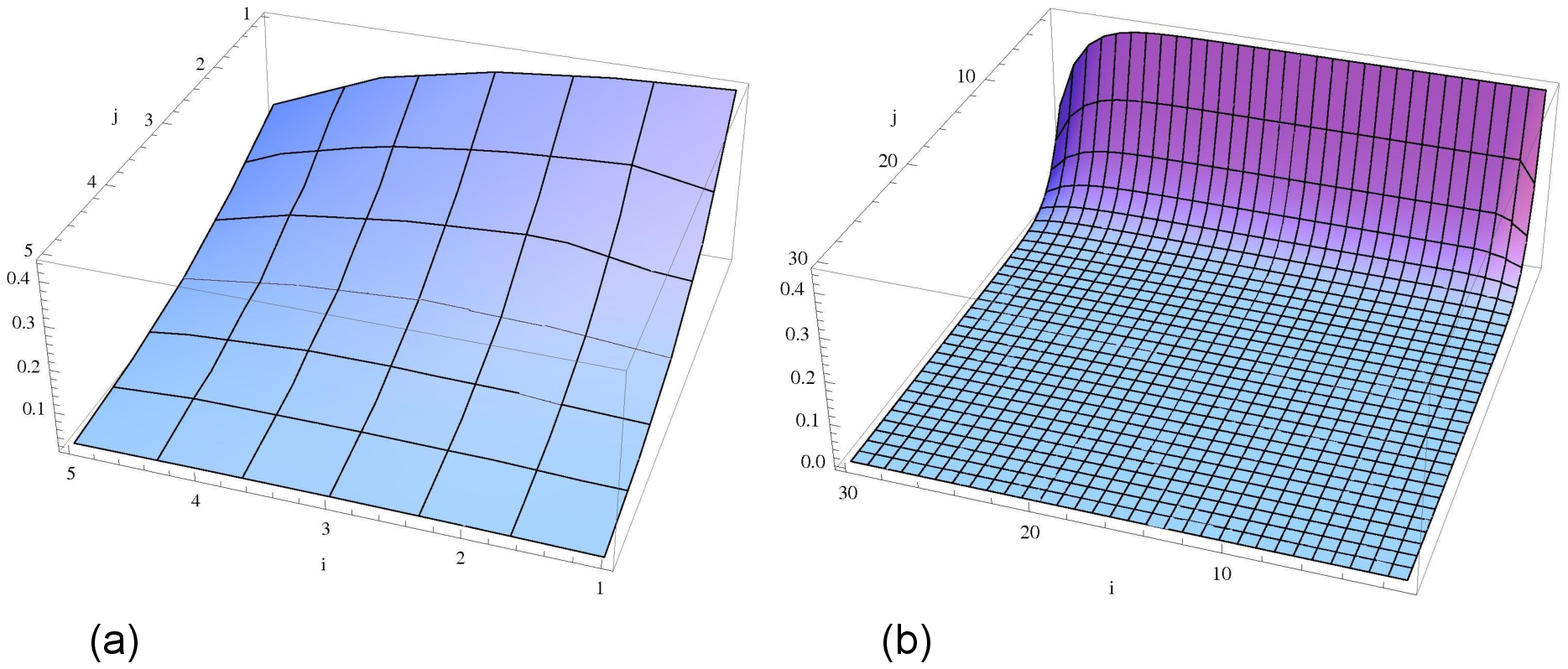,width=10cm} \epsfig{file=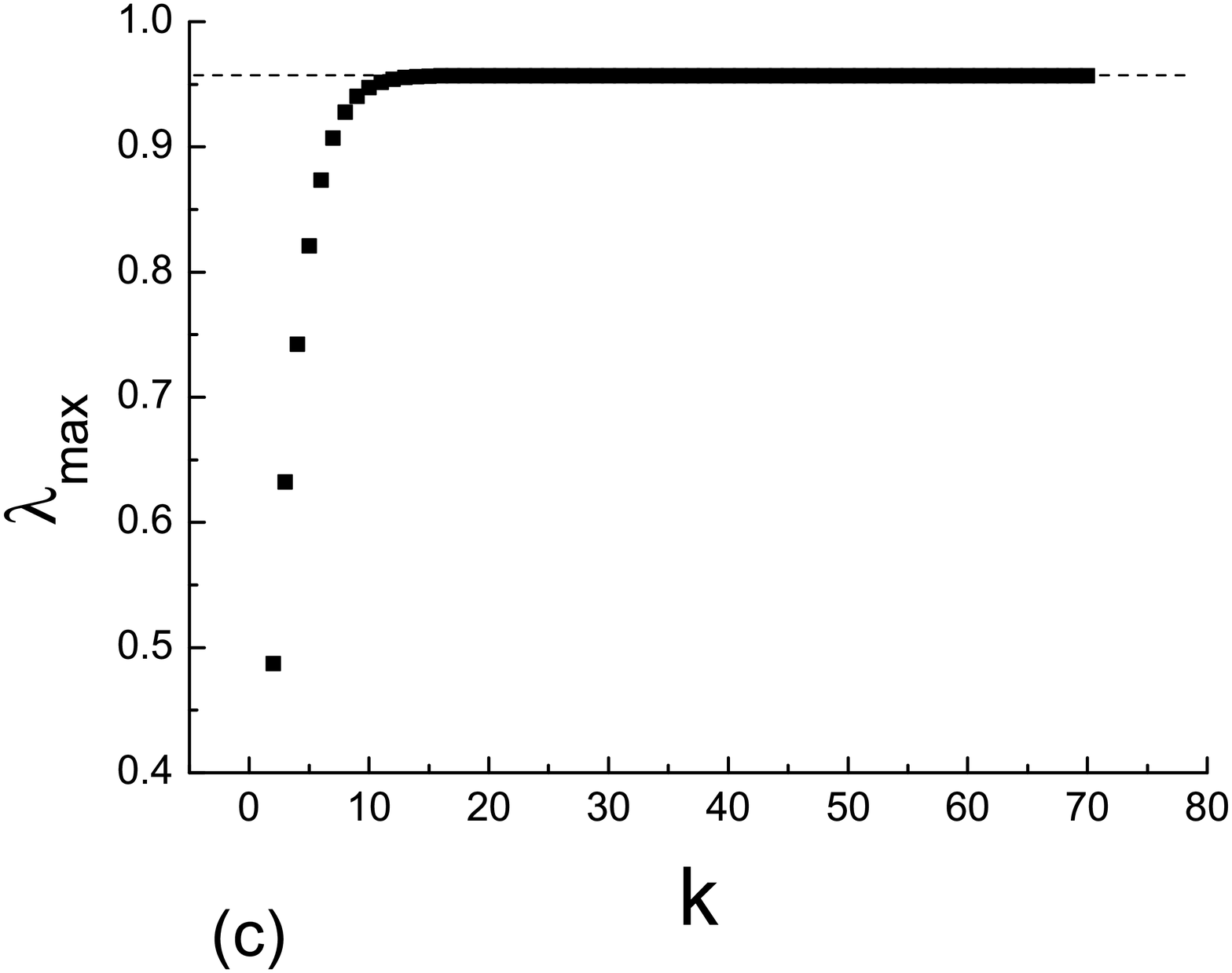,width=5cm}
\caption{Visualization of the matrix $B$ as a 3D plot of $B_{i,j}$ over the base $(i,j)$ for: (a) $k=5$,
(b) $k=30$; (c) Dependence on the maximal eigenvalue of the matrix $B$ (see \eq{eq:04}) of size
$k\times k$ on $k$. The limiting value of $\lambda_{\max}$ tends to $\approx0.957$ as
$k\to\infty$.}
\label{fig:03}
\end{figure}

In the \fig{fig:03}a,b we have plotted the matrix elements of $B$ for $k=5,30$ in form of a 3D
relief $B_{i,j}$ over the base plane $(i,j)$, where $(i,j)=1,...,k$. Computing numerically the
maximal eigenvalues of the true matrices $B$ defined in \eq{eq:04}, for different $k$, we get a
plot shown in the \fig{fig:03}c.

Extrapolating the data shown in \fig{fig:03} to $k\to\infty$, we obtain
\be
\lambda_{\rm max} \approx 0.957
\label{eq:09}
\ee
meaning that the largest eigenvalue, $\lambda_{\max}$ of the matrix $B$ and, consequently, of the
initial Kramers matrix $G$, is bounded from above by some numeric value independent on the matrix
size (and without any logarithmic corrections).

It is noteworthy that a rough estimate of the eigenvalue $\lambda_{\max}$ can be done using the
following simple analytical argument. All matrix elements $B_{i,j}$ are positive and
$$
B_{i,j} =\left\{\begin{array}{ll}b_{i,j} & \quad j\ge i \medskip \\
b_{j,i}\, 2^{i-j} & \quad j<i
\end{array} \right.
$$
Taking into account the explicit form of matrix elements $b_{i,j}$ for $j\ge i$ (see \eq{eq:04}), we
can approximate each element $B_{i,j}$ for any $(i,j)$ by $b_{i,j}$. This approximation gives exact
value of $B_{i,j}$ for $j\ge i$, and provides an upper estimate of elements $B_{i,j}$ for $j<i$.

The maximal eigenvalue, $\bar{\lambda}_{\rm max}$, of the $k\times k$ matrix $\bar{B}$ with entries
$\{b_{i,j}\}$ can be computed exactly. Eq.\eq{eq:04} means that the element $b_{i,j}$ can be
factorized as
\be
b_{i,j} = f_i\, g_j; \quad f_i = 2^{-1 - 2 i}(2^{3 + k} - 3\times 2^{2 + i + k} + 3\times 2^{2i});
\quad g_j=\frac{2^j - 2^{1 + k}}{(3\times 2^k-2)^2}
\label{eq:factor}
\ee
Now, the largest eigenvalue, $\bar{\lambda}_{\rm max}$, reads
\be
\bar{\lambda}_{\max} = \sum_{m=1}^k f_m\, g_m = \frac{7\times 2^{2 + 2 k} - 27 k\times 2^k -
15\times 2^k - 13}{3(3\times 2^k-2)^2}
\label{eq:07}
\ee
For $k\to\infty$ on gets from \eq{eq:07}
\be
\bar{\lambda}_{\max}\Big|_{k\to\infty} =\frac{28}{27} \approx 1.037
\label{eq:08}
\ee
which is a pretty good approximation for numerically found true limiting value
$\lambda_{\max}\approx 0.957$.

\subsection{Sparse regular trees}

We can generalize our approach to ``sparse'' regular trees, in which branchings are connected by
linear subchains of $s$ links. By definition, for $s=1$ we return to the former case of branching
at every node of the tree. Particular example of a dendrimer of $N=19$ nodes with 3-branching
vertices separated by 2-link subchains (i.e. $s=2$), is shown in the \fig{fig:01}b.

To construct the Kramers adjacency matrix, $G^{(s)}$, for such a tree, one proceeds as above, using
the descent diagram representation, where we first enumerate sequentially links in each linear
subchain, and then pass to the next subchain in the same level (compare \fig{fig:01}a and
\fig{fig:01}b). The matrix $G^{(s)}$ constructed in such a way, highly resembles the former one,
$G$, where each matrix element $G_{x,y}$ is replaced by a $s\times s$--block (a $2\times 2$ block
for a particular example of \fig{fig:01}b).

The largest eigenvalue, $\lambda_{\max}^{\rm tree}(s)$, of the composite adjacency matrix,
$G^{(s)}$, can be estimated as
\be
\lambda_{\max}^{\rm tree}(s) \approx \lambda_{\max}^{\rm lin}(s)\; \lambda_{\max}^{\rm tree}
\label{eq:10}
\ee
where $\lambda_{\max}^{\rm lin}(s)$ is the maximal eigenvalue of a Kramers matrix for a linear
chain of $s$ links and $\lambda_{\max}$ is given by \eq{eq:09}. The value of $\lambda_{\max}^{\rm
lin}(s)$ reads:
\be
\lambda_{\max}^{\rm lin}(s) = \frac{s+1}{4s^2} \sin^{-2}\frac{\pi}{2(s+1)}; \quad
\lambda_{\max}^{\rm lin}(s) = \left\{\begin{array}{l} \lambda_{\max}^{\rm lin}(s=1)=1 \medskip \\
\lambda_{\max}^{\rm lin}(s\gg 1) =\frac{s}{\pi^2}
\end{array} \right.
\label{eq:11}
\ee
As it follows from \eq{eq:09}, the value $\lambda_{\max}$ can be approximated by some constant,
$c$, of order of unity. Thus, we can estimate $\lambda_{\max}^{\rm tree}(s)$ for all regular sparse
trees as
\be
\lambda_{\max}^{\rm tree}(s) \approx c \lambda_{\max}^{\rm lin}(s) = \bar{c}\, s
\label{eq:12}
\ee
where $\bar{c}$ absorbs all numerical constants.

\section{Discussion}

Let us now summarize what we learned about trees. To begin with, recall that for a linear polymer the
whole spectrum of the Kramers matrix is known \cite{Fixman_1962} (see also small corrections in
\cite{Fixman_correction_1} and also \cite{Fixman_correction_2}); in particular
$\lambda_{\max}=\frac{N}{\pi^2}$. Since spanning distance in this case is just $L=N$, we have
$\lambda_{\max} = \frac{L}{\pi^2} \sim L$.

Next, for the perfect 3-branching dendrimer we have established above that $\lambda_{\max} \approx
1$. This has to be compared with average spanning distance of this tree.  For dendrimer with $k$
generations and, accordingly, $3 \times 2^k$ ends, the sum of distances along a backbone from one
end to all other ends is equal to $2 + 2^{k+1}(3k+1)$. Hence, $L = \frac{1 + 2^{k}(3k+1)}{3 \times
2^{k-1}}$, and at large $k$ this asymptotically tends to $L \simeq 2k$.  Since $N = 3 \times
2^{k+1} -2 $, (or, equivalently, $k \simeq \mathrm{log}_2 \frac{N}{3} - 1 \simeq \log_2 N$), we
have $L \simeq 2 \log_2 N$. Therefore, we can say that $\lambda_{\max} \approx \frac{L}{2} \log_2
N$.

Thus, for a regular tree, $\lambda_{\max}$ is related to the average spanning distance $L$ via a
factor of order $\ln N$.  Let us show now that this is the worst case, and for all other trees the
estimate $\lambda_{\max} \sim L$ is even more accurate.

Sparse regular trees provide in this sense a good insight, as they smoothly interpolate between
regular tress and linear chains.  As we have seen, in this case $\lambda_{\max} \approx s$. On the
other hand, in this case $N= 3s \left(2^{k+1}-1 \right) +1$, or $k = \log_2 \left[
\frac{N+3s-1}{3s}\right]$; at large $N$ this is asymptotic to $k \simeq \log_2
\left[\frac{N}{s}\right]$. At the same time, $L \simeq 2 s k $ or $L \simeq 2 s  \log_2 \left[
\frac{N}{s} \right]$.  Thus,
$$
\lambda_{\max} \approx \frac{L}{2} \log_2 \left[ \frac{N}{s} \right]
$$
We see that the factor relating $\lambda_{\max}$ to mean spanning distance $L$ smoothly changes
between about $\ln N$ and about $1$ as $s$ changes from $1$ to $N$.

In fact, it is clear physically that a regularly branching tree is the most compact and most
difficult to swell of all quenched branched polymers, while linear polymer is the least compact and
the easiest to swell.  Since we have rigorously analyzed both of these extremes, we conclude that
the expression (\ref{eq:Kramers2}), which establishes conformational entropy up to logarithmic
corrections, implies also the validity of the scaling estimate of entropy (\ref{F_in_terms_of_L})
in terms of average spanning diameter.

\acknowledgements This work has been started in Paris, where AYG was on a long term visit.  AYG
acknowledges the hospitality of both the Curie Institute and ESPCI. Both authors are very grateful
to R. Everaers, Y. Fyodorov, and J. Smrek for useful discussions. The work of SKN was partially
supported by DIONICOS European project.

\begin{appendix}

\section{Generalized Kramers theorem}
\label{sec:Kramers}

Here, we compute the gyration radius probability distribution for a Gaussian tree. The approach
follows the works of M. Fixman \cite{Fixman_1962, Fixman_correction_1, Fixman_correction_2}, and
\cite{Moore_gyration}, where it was generalized for Gaussian rings.

Suppose for simplicity that the branching number is $p=3$, and there are only branchings and ends,
namely, $n$ three-valent vertices and $n+2$ ends, totally $N=2n+2$ monomers. If ${\bf r}_{i}$ are
position vectors of these ``monomers,'' then the gyration radius reads
\be
R^{2} = \frac{1}{2N^{2}} \sum_{i,j} ({\bf r}_{i}-{\bf r}_{j})^{2}.
\ee
On the tree, every ${\bf r}_{i}-{\bf r}_{j}$ is uniquely represented as the sum of the set of bond
vectors ${\bf a}_{i}={\bf r}_{i^{\prime}}-{\bf r}_{i}$, where $i$ and $i^{\prime}$ are monomers
connected by bond $k$. Accordingly, gyration radius can be represented as
\be
R^{2} = \sum_{x,y} G_{x,y} {\bf a}_{x}{\bf a}_{y},
\label{RgG}
\ee
where the indices $x$ and $y$ label bonds (unlike $i$, $i^{\prime}$ and $j$ above, which label
vertices or monomers), and $G_{x,y}$ is an Kramers $(N-1) \times (N-1)$ matrix, illustrated in the
\fig{fig:Gmatrix}.

For a Gaussian tree we can do more and find the probability distribution of $R^2$, assuming each
bond has the probability distribution $\sim e^{-3{\bf a}^{2}/2a^{2}}$. The characteristic function
of $R^{2}$ reads
\be
\Phi(s) = \la e^{i s R^{2}}\ra = A \int d^{3}\left\{{\bf a}\right\} e^{-\sum_{k} 3{\bf
a}_{k}^{2}/(2a^{2}) + i s \sum_{x,y} G_{x,y}{\bf a}_{x}{\bf a}_{y}} ,
\ee
where the explicit expression for normalization factor $A$ is dropped for brevity. Rotating now the
coordinate system in this $N-1$  dimensional space of $\left\{{\bf a}\right\}$ to the basis of
eigenvectors $\{{\bf b}\}$ of matrix $G$, we obtain
\be
\Phi(s) = A\int d^{3}\{{\bf b}\} e^{-\sum_{p} {\bf b}_{p}^{2} \left( \frac{3}{2a^{2}} - i s
\lambda_{p} \right)} = \prod_{p=1}^{N-1} \left(1- \frac{2 a^2  i s}{3 } \lambda_{p} \right)^{-3/2},
\ee
where $\lambda_{p}$ are eigenvalues, and normalization factor $A$ is reconstructed from the
condition $\left. \Phi(s) \right|_{s=0} = 1$. From here, finding the probability distribution of
$R^{2}$ is a matter of inverse Fourier transform of $\Phi(s)$
\be
P(R^{2}) = \frac{1}{2\pi} \int ds\; e^{-i s R^{2}- \frac{3}{2}\sum_{p}\ln \left( 1- \frac{2 a^2 i
s}{3} \lambda_{p} \right)} \ .
\ee

Since we are interested in the behavior of $P(R^{2})$ at large $R$, the asymptotics is controlled
by the singularity of $\Phi(s)$ closest to the origin in the complex $s$--plane, that corresponds
to the largest eigenvalue $\max_p[\lambda_{p}]=\lambda_{\max}$. In the vicinity of this singularity
there is a saddle point which dominates the integral, and we can evaluate inverse Fourier transform
integral by steepest descent. The equation for saddle point location reads $R^{2}\simeq \frac{a^2
\lambda_{\mathrm{max}}}{1-i s (2 a^2 / 3) \lambda_{\mathrm{max}} }$, or $s= i \left( \frac{3}{2
R^2} - \frac{3}{2 a^2 \lambda_{\max}}\right)$. Thus, the result of saddle point integration (up to
logarithmic corrections) is
\be
P(R^{2})|_{R\to\infty} \sim e^{-\frac{3 R^{2}}{2 a^2 \lambda_{\max}} +
\frac{3}{2}\ln\frac{R^{2}}{\lambda_{\max}}} \sim e^{-3 R^{2}/2 a^2 \lambda_{\max}} \ ,
\ee
yielding the expected generalization of Kramers theorem (\ref{eq:Kramers2}).

\end{appendix}

\end{document}